\documentclass[a4paper,11pt]{article}
\usepackage[reqno,fleqn]{amsmath}
\usepackage{color,graphicx}
\usepackage[T1]{fontenc}
\author{Behnam.Mohammadi$^a$\footnote{be.mohammadi@urmia.ac.ir}, Mahdi Lotfizadeh$^b$\footnote{m.lotfizadeh@urmia.ac.ir}\\
Department of Physics, Urmia University, Urmia, Iran}
\title{Evaluation of the $B^+_c\rightarrow D^0K^+$ decay by the factorization approaches and applying the effects of the final state
interaction}
\begin{document}
\maketitle
\begin{abstract}
In this paper the decay of $B^+_c$ meson, consisting of two b and
c heavy quarks, into the $D^0$ and $K^+$ mesons is studied. Given
that the experimental branching ratio for this decay is within the
range of $3.72\times 10^{-5}$ to $11.16\times10^{-5}$ and in our
estimating the theoretical result by using the QCD factorization
approaches is $10^2$ times less than experimental one (we have
obtained $1.41\times10^{-7}$), it is decided to calculate the
theoretical branching ratio by applying the final state
interaction (FSI) through the T and cross section channels. In
this process, before the $B^+_c$ meson decays into two final state
mesons of $D^0K^+$, it first decays into two intermediate mesons
like $J/\psi D^{*+}_s$, then these two mesons transformed into two
final mesons by exchanging another meson like $D^0$. The FSI
effects are very sensitive to the changes in the phenomenological
parameter that appear in the form factor relation, as in most
calculation changing two units in this parameter, makes the final
result multiply in the branching ratio, therefor the decision to
use FSI is not unexpected. In this study there are nineteen
intermediate states in which the contribution of each one is
calculated and summed in the final amplitude. Therefore, the
numerical value of the branching ratio of $B^+_c\rightarrow
D^0K^+$ decay is obtained by calculating the FSI effects from
$1.17\times10^{-5}$ to $11.65\times10^{-5}$ which is consistent
with the experimental result.
\end{abstract}

\section{Introduction}

The discovery of $B^+_c$ meson was first reported by CDF
collaboration at the Fermi Lab in 1998, during the process of the
$B^+_c\rightarrow J/\psi \ell^+\nu_\ell$ decay \cite{CDF1}. After
that, in 2008 the $B^+_c\rightarrow J/\psi\pi^+$ decay was
observed by CDF and $D^0$ collaborations with $8\sigma$
\cite{CDF2} and $5\sigma$ \cite{V.Ab}, respectively. Also the
$B^+_c\rightarrow J/\psi\pi^+$ decay in 2013 was observed by LHCb
collaboration in proton-proton collision with center of mass
energy 7 TeV. The $B^+_c$ meson is the only meson that has been
observed until now that has two heavy quarks with different
flavors. Both of these quarks have a strong desire for decay. This
meson can not be destroyed to produce a gluon, it can only be
decayed by weak interaction. For this reason the meson $B^+_c$ is
a good case to study the mechanism of weak decay of heavy flavors
and evaluation of quark-flavor mixing in the standard model. The
$B^+_c$ meson has many weak decay channels,
all of them can be classified into three different classes:\\
1- The decay of b-quark into two c or u-quarks, which at this step
another c-quark is added as a spectator to the collection.\\
2- The decay of c-quark into two s or d-quarks which a b-quark
used as a spectator.\\
3- Weak annihilation decay channel.\\
For the class 1, the $|V_{ub}|$ and $|V_{cb}|$ matrix elements of
CKM matrix elements are of interest.\\
For the class 2, in which the heavy mesons $B_s$ or $B_d$ are in
the final state can have significant effects for the c-quark decay
in the phase space. In this class, $|V_{cs}|$ and $|V_{cd}|$ of
CKM matrix elements are of interest which in comparison with the
elements $|V_{ub}|$ and $|V_{cb}|$ used in class 1 have much
larger values.\\
In the class 3, the weak annihilation decay of $B^+_c$ meson has
significant amount compared with the $B^+_u$ decay so that the
ratio of $|V_{cb}|^2$ to $|V_{ub}|^2$ in weak annihilation decay
of $B^+_c$ and $B^+_u$, is approximately 100, in fact, we have
$|V_{cb}|^2/|V_{ub}|^2\sim 100$. This means that unlike $B^+_u$
decay, which can be ignored, the annihilation decay step has a
significant value. In the study of $B^+_c$ meson decay, all three
classes are very important. Unlike $B^+_u$ and $B^0_d$ mesons
decay, more than $70\%$ of the $B^+_c$ meson decays occurs by
c-quark decay which in this context the transition $c\rightarrow
s$ with $B^+_c\rightarrow B^0_s\pi^+$ has been observed. The
b-quark decay has only near to $20\%$ of $B^+_c$ meson decays. In
the case that there is no c-quark in the final state, the
$\bar{b}c\rightarrow W^+\rightarrow \bar{q}q$ annihilation
amplitudes are only $10\%$ of the total $B^+_c$ meson decays. The
$B^+_c\rightarrow D^0K^+$ branching ratio was calculated using
perturbative quantum chromodynamics (pQCD) method before it was
observed in the experiment, which has been obtained the value of
$6.60\times10^{-5}$ \cite{J.Zh}. It has also been solved using the
factorization approaches which is that result of
$1.34\times10^{-7}$ \cite{H.f.F} which is in good agreement with
what we have achieved in this way. Until the decay
$B^+_c\rightarrow D^0K^+$ has been observed by LHCb collaboration,
they have obtained $B^+_c$ production compared to $B^+_u$ as
\cite{LHCb1}: {\setlength\arraycolsep{.75pt}
\begin{eqnarray}
\frac{f_c}{f_u}\mathcal{B}(B^+_c\rightarrow
D^0K^+)=(9.3^{+2.8}_{-2.5}\pm0.6)\times10^{-7},
\end{eqnarray}}
in which, the ratio $f_c/f_u$ is an unknown value obtained by
evaluating two decays $B^+_c\rightarrow J/\psi \pi^+$ \cite{D.Eb,
C.H.C} and $B^+_c\rightarrow J/\psi K^+$ \cite{M.Ta} in the range
of 0.004 to 0.012. In this case, the branching ratio that LHCb
collaboration have obtained is: {\setlength\arraycolsep{.75pt}
\begin{eqnarray}
\mathcal{B}(B^+_c\rightarrow
D^0K^+)=(3.72^{+1.12}_{-1.00}\pm0.24)\times10^{-5}\sim
(11.16^{+3.36}_{-3.00}\pm0.72)\times10^{-5}.
\end{eqnarray}}
It is clear that the factorization approach is about 100 times
smaller than experience, while the pQCD result is within the
experimental range. In \cite{H.f.F}, unfactorizable contributions
are not considered, which reflects the fact that the heavy
observed gluon contributions in strong interactions is neglected.
By doing this, it will not matter how other parameters like strong
phase and phenomenological parameter can be adjusted. The
phenomenological parameter is a parameter that appears in the FSI
form factors that increase strong interaction share. The final
results are very sensitive to this parameter so that, the range of
final results with a little change in phenomenological parameter
changes dramatically.\\
In \cite{LHCb1}, it is explicitly stated that this decay is
expected to continue with penguin and weak annihilation amplitudes
but, as we know, the main contribution of the final amplitude lies
in the tree amplitudes. In fact, if we remove the tree amplitude
share from what was done in \cite{H.f.F}, the result will be
$10^3$ times smaller than the experience. By entering
unfactorizable share, one can not compensate $10^3$ times the
smaller of the result. It seems, it is needed a model, a method or
applying natural effects to compensate this major difference, we
enter FSI effects. As it was said, the experimental range of
$B^+_c\rightarrow D^0K^+$ decay is within the range of $3.72\times
10^{-5}$ to $11.16\times10^{-5}$, that is, a relatively large
range, this is why we decided to re-calculate the branching ratio
of this decay by entering FSI effects. In the previous works
\cite{B.M1,B.M2,B.M3}, we have seen that the calculation of
intermediate state effects is very sensitive to phenomenological
parameter. In some cases, by changing the two units in the value
of this parameter, the final result is changed to several times.
In the $B^+_c\rightarrow D^0K^+$ decay, since the limit of the
experimental result is approximately three times, the
phenomenological parameter change can cover the range of
experimental results. In this paper, we are talking about the fact
that during the $B^+_c\rightarrow D^0K^+$ decay some middle
particles are produced, in the way that before $D^0$ and $K^+$
particles occur in the final state the middle particles are formed
which have been converted into final mesons by exchanging another
particle. The process of producing these middle and exchanged
particles is determined through Feynman diagrams.\\
For FSI quark model, the Feynman graphs are presented in two
types: the first one is the T-channel and the second one is the
cross section-channel. In the T-channel, two final mesons of $D^0$
and $K^+$ share one quark and one anti-quark with the same flavor
(u). The intermediate mesons are produced by sharing c, d and s
quarks. In this case, the intermediate state mesons $J/\psi
D^{+(*)}_s$ (both $J/\psi D^+_s$ and $J/\psi D^{*+}_s$),
$D^{+(*)}K^{0(*)}$ and $D^{+(*)}_s\phi$ can be produced with
$D^0$, $\pi^-$ and $K^-$ exchange mesons, respectively. In the
cross section channel, two final mesons exchange one non-flavored
quark with intermediate mesons crosswisely. For the
$B^+_c\rightarrow D^0K^+$ decay, the two final state mesons $D^0$
and $K^+$ exchange c and u quarks with intermediate mesons,
crosswisely. In this case, $D^{+(*)}p$ mesons in which p can be
$\pi^0, \rho^0, \eta$ and $\omega$ in the intermediate and $D^0$
meson as the exchange meson can be presented in this process.
Also, two final mesons $D^0$ and $K^+$ each of them can exchanges
one anti-particle $\bar{u}$ and $\bar{s}$, respectively, with
intermediate state mesons, crosswisely. Then, intermediate mesons
will be the same as the previous mesons $D^{+(*)}p$ (which p-types
have already been identified), while exchanged meson in this
process is $K^-$. In general, the $B^+_c\rightarrow D^0K^+$ decay
is transformed into following decays using FSI effects on the
T-channel: {\setlength\arraycolsep{.75pt}
\begin{eqnarray}
B^+_c&\rightarrow& J/\psi D^{+(*)}_s\rightarrow D^0K^+\;\;
\rm{exchang\;meson\;is}\;D^0,\nonumber\\
B^+_c&\rightarrow& D^{+(*)}K^{0(*)}\rightarrow D^0K^+\;\;
\rm{exchang\;meson\;is}\;\pi^-,\nonumber\\
B^+_c&\rightarrow& D^{+(*)}_s\phi\rightarrow D^0K^+\;\;
\rm{exchang\;meson\;is}\;K^-,
\end{eqnarray}}
and transformed into the following decays in the cross
section-channel: {\setlength\arraycolsep{.75pt}
\begin{eqnarray}
B^+_c&\rightarrow& D^{+(*)}_sP\rightarrow D^0K^+\;\;
\rm{exchang\;meson\;is}\;D^0,\nonumber\\
B^+_c&\rightarrow& D^{+(*)}_sP\rightarrow D^0K^+\;\;
\rm{exchang\;meson\;is}\;K^-.
\end{eqnarray}}
As an example, for the intermediate state $B^+_c\rightarrow J/\psi
D^{+(*)}_s$, it can be said that before the two mesons $D^0$,
$K^+$ are produced in the final state, two mesons $J/\psi$ and
$D^{+(*)}_s$ which are produced in the intermediate state are
transformed to the final state mesons by exchanging $D^0$ meson.
To calculate the total amplitude of the $B^+_c\rightarrow D^0K^+$
decay, the five channels listed above (three T-channels and two
cross section-channels) which are calculated using FSI method,
should be added. In the calculations, we also need the individual
amplitudes of the intermediate states. So, in the next section, we
calculate the intermediate state amplitudes.

\section{Short distance processes}
\subsection{Amplitude of $B^+_c\rightarrow D^0K^+$ decay by using the QCD factorization approaches}

\begin{figure}[t]
\centering \includegraphics[scale=0.9]{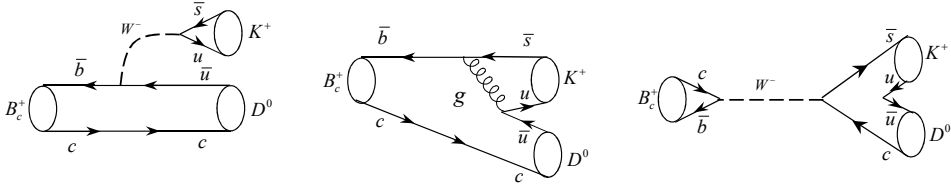}
\caption{\label{fig:b1}Diagrams for $B^+_c\rightarrow D^0K^+$
decay.}
\end{figure}

A detailed discussion of the QCD factorization approaches can be
found in \cite{A.Al1,M.Be1}. Factorization is a property of the
heavy-quark limit, in which we assume that the b quark mass is
parametrically large. The QCD factorization formalism allows us to
compute systematically the matrix elements of the effective weak
Hamiltonian in the heavy-quark limit for certain two-body final
states $D^0K^+$. In this section, we obtain the amplitude of
$B^+_c\rightarrow D^0K^+$ decay using QCD factorization method.
Under the factorization approach, there are color-allowed tree and
suppressed penguin diagrams to $B^+_c\rightarrow D^0K^+$ decay. We
adopt leading order Wilson coefficients at the scale $m_b$ for QCD
factorization approach. The diagrams describing this decay are
shown in Fig. \ref{fig:b1}.
According to the QCD factorization, the amplitude of
$B^+_c\rightarrow D^0K^+$ decay is given by
{\setlength\arraycolsep{.75pt}
\begin{eqnarray}
M_{QCDF}(B^+_c\rightarrow
D^0K^+)=\frac{G_F}{\sqrt{2}}f_{K^+}\Big[&&(m_{B^+_c}^2-m_{D^0}^2)F^{B^+_c\rightarrow
D^0}(m_{K^+}^2)(a_1V^*_{ub}V{us}+a_4V^*_{tb}V_{ts})\nonumber\\
&&+f_{B^+_c}f_{D^0}b_2V^*_{cb}V_{cs}\Big],
\end{eqnarray}}
where $a_1$ and $a_4$ correspond to the current-current tree and
penguin, and $b_2$ corresponds to the current-current annihilation
coefficients that are given by {\setlength\arraycolsep{.75pt}
\begin{eqnarray}
a_1&=&c_1+\frac{c_2}{N_c},\quad a_4=c_4+\frac{c_3}{N_c},\nonumber\\
b_2&=&\frac{C_{F}}{N_{c}^{2}}c_2A_1^i,
\end{eqnarray}}
$c_i$ are the Wilson coefficients, $N_c=3$ is the color number and
{\setlength\arraycolsep{.75pt}
\begin{eqnarray}
A_1^i&=&2\pi\alpha_{s}[9(X_{A}-4+\frac{\pi^{2}}{3})+r_{\chi}^{D^0}r_{\chi}^{K^+}X_{A}^{2}],\nonumber\\
C_F&=&\frac{N_c^2-1}{2N_c},
\end{eqnarray}}
where $r_{\chi}^{D^0}=1.85$ and $r_{\chi}^{K^+}1.14$. For the
running coupling constant, at two loop order (NLO) the solution of
the renormalizaton group equation can always be written in the
form
\begin{eqnarray}
\alpha_s(\mu)=\frac{4\pi}{\beta_{0}ln\frac{\mu^2}{\Lambda_{QCD}^2}}[1-\frac{\beta_1}{\beta_{0}^2}
\frac{ln(ln\frac{\mu^2}{\Lambda_{QCD}^2})}{ln\frac{\mu^2}{\Lambda_{QCD}^2}}],
\end{eqnarray}
here
\begin{eqnarray}
\beta_0=\frac{11N_c-2n_f}{3},\;\beta_1=\frac{34N_c^2}{3}-\frac{10N_cn_f}{3}-2C_Fn_f,
\end{eqnarray}
and running $\alpha_s(\mu)$ evaluated with $n_f=5$. There are
large theoretical uncertainties related to the modeling of power
corrections corresponding to weak annihilation effects, we
parameterize these effects in terms of the divergent integrals
$X_A$ (weak annihilation)
\begin{eqnarray}
X_{A}=(1+\rho
e^{i\phi})\ln{\frac{m_{B^+_c}}{\Lambda_h}},\label{eq10}
\end{eqnarray}
where, can be obtained by using $\rho=0.5$, $\phi=-55^\circ$ and
$\Lambda_h=0.5Gev$.

\subsection{Decay amplitudes of intermediate states}

\begin{figure}[t]
\centering \includegraphics[scale=0.9]{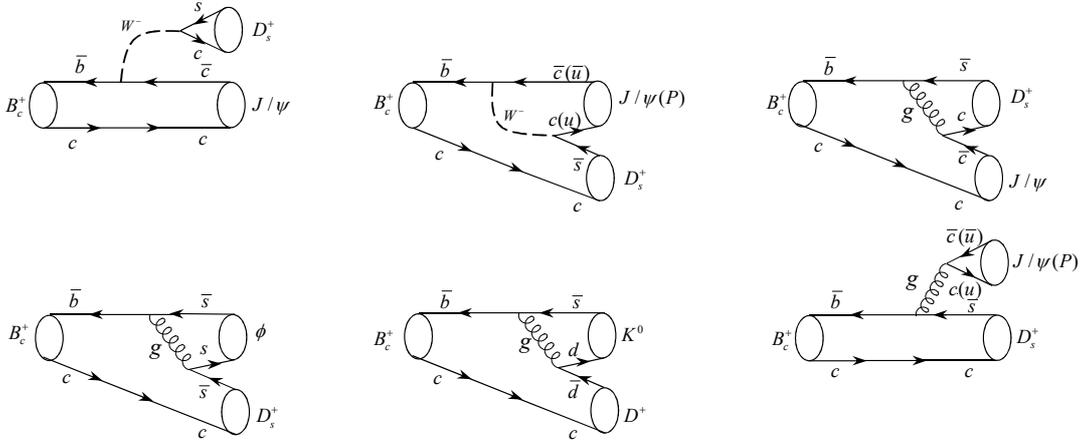}
\caption{\label{fig:b2}Feynman diagrams for intermediate states
decays.}
\end{figure}

Each decay in the intermediate states is a two body decay of
$B^+_c$ meson that until now the amplitude of such decays has been
achieved in many ways: naive factorization, QCD factorization,
improved factorization and using QCD perturbation. In present
section, we
use QCD factorization to obtain intermediate state amplitudes.\\
The first intermediate state decay is: $B^+_c\rightarrow J/\psi
D^{+(*)}_s$. Usually, in FSI, the dominant contribution of
intermediate amplitudes is considered, however, here we consider
all of them. The decay $B^+_c\rightarrow J/\psi D^+_s$ happens
both through $\bar{b}\rightarrow\bar{c}$ and
$\bar{b}\rightarrow\bar{s}$ transitions. The tree transition of
$\bar{b}\rightarrow\bar{c}$ concludes two contributions of
Wilson's coefficients: $a_1$ and $a_2$. The matrix elements of
both contributions are $V_{cb}^*V_{cs}$. For $a_1$ contribution,
the mesons of $J/\psi$ and $D^+_s$ place in form factor and decay
constant, respectively. But, in $a_2$ contribution, conversely.\\
The penguin transition $\bar{b}\rightarrow\bar{s}$ has $a_3$ and
$a_4$ Wilson contributions. In $a_3$ contribution, mesons of
$D^+_s$ and $J/\psi$ place in the form factor and decay constant,
respectively. But, in the $a_4$ coefficient the rule is the
opposite of this. Corresponded matrix elements are
$V_{tb}^*V_{ts}$.\\
The second and third intermediate state decays are
$B^+_c\rightarrow D^{+(*)}_s\phi$ and $B^+_c\rightarrow
D^{+(*)}K^{0(*)}$. These decays only have penguin transition. In
fact, they have just $a_4$ contribution in which mesons of $D^+_s$
and $D^+$ place in form factor and mesons of $\phi$ and $K^0$
place in decay constant. Their CKM matrix elements are
$V_{tb}^*V_{ts}$.\\
The fourth and fifth intermediate state decays are the decays of
$B^+_c\rightarrow D^{+(*)}_s P$ with $P=\pi^0, \rho^0, \omega,
\eta$ which include the $a_2$ and $a_3$ contributions of tree
transition $\bar{b}\rightarrow\bar{u}$ and penguin transition
$\bar{b}\rightarrow\bar{s}$, respectively. The corresponding
matrix elements are $V_{ub}^*V_{us}$ and $V_{tb}^*V_{ts}$,
respectively. The appearance of the five decays mentioned above
can be seen in the Fig. \ref{fig:b2}.
In all these decays, there were contributions of Wilson. These
contributions are derived from the combination of Wilson
coefficients. If these coefficients are used in the normal from,
the factorization is called naive factorization, while, using
Wilson effective coefficients, it is called QCD factorization.
Wilson tree and penguin contributions get from
$a_{1,2}=c_{1,2}+c_{2,1}/3$ and $a_{3,4}=c_{3,4}+c_{4,3}/3$,
respectively. The $c_i$ are normal Wilson coefficients, these
become Wilson effective coefficients if the vertex correction and
hard gluon scattering are taken into account ($c_i\rightarrow
c_i^{eff}$). In all of these decays, we used terms like: tree
transition, penguin transition, decay constant and form factor.
The form factors and decay constants for pseudo scalar and vector
mesons can be written respectively as \cite{A.Al1}:
{\setlength\arraycolsep{.75pt}
\begin{eqnarray}
\langle
P(p')|V_\mu|B(p)\rangle&=&[(p+p')_\mu-\frac{m_B^2-m_P^2}{q^2}q_\mu]F_1(q^2)+\frac{m_B^2-m_P^2}{q^2}q_\mu
F_0(q^2)\nonumber\\
\langle 0|A_\mu|P(q)\rangle&=&if_pq_\mu\nonumber\\
\langle V(\epsilon,p')|V_\mu-A_\mu|B(p)\rangle&=&(\epsilon^*\cdot
q)\frac{2m_V}{q^2}q_\mu A_0(q^2)+(m_B+m_V)
[\epsilon^*_\mu-\frac{\epsilon^*\cdot q}{q^2}q_\mu]A_1(q^2)\nonumber\\
&&-\frac{\epsilon^*\cdot q}{m_B+m_V}[(p+p')_\mu -\frac{m_B^2-m_V^2}{q^2}q_\mu]A_2(q^2)\nonumber\\
\langle 0|V_\mu|V(\epsilon,q)\rangle&=&if_Vm_V\epsilon_\mu,
\end{eqnarray}}
where p and v are pseudo scalar and vector mesons, respectively.
The q parameter is the four-momentum of propagator the square of
which is $q^2=m_B^2+m_p^2-2m_Bp^0_p$. So, amplitudes of
intermediate decays take the following form:
{\setlength\arraycolsep{.75pt}
\begin{eqnarray}
M(B^+_c\rightarrow J/\psi
D^+_s)&=&i\sqrt{2}G_Fm_{J/\psi}(\epsilon_{J/\psi}\cdot
p_B)\{[a_1f_{D^+_s}A_0^{B^+_c\rightarrow
J/\psi}(m^2_{D^+_s})+a_2f_{J/\psi}F_1^{B^+_c\rightarrow
D^+_s}(m^2_{J/\psi})]\nonumber\\
&&\times V^*_{cb}V_{cs}+[a_3f_{J/\psi}F_1^{B^+_c\rightarrow
D^+_s}(m^2_{J/\psi})+a_4f_{D^+_s}A_0^{B^+_c\rightarrow
J/\psi}(m^2_{D^+_s})]V^*_{tb}V_{ts}\}\nonumber\\
M(B^+_c\rightarrow J/\psi
D^{*+}_s)&=&i\frac{G_F}{\sqrt{2}}\{f_{D^{*+}_s}m_{D^{*+}_s}[(\epsilon_{J/\psi}\cdot\epsilon_{D^{*+}_s})(m_{B^+_c}+m_{J/\psi})
A_1^{B^+_c\rightarrow J/\psi}(m^2_{D^{*+}_s})\nonumber\\
&&-(\epsilon_{J/\psi}\cdot p_{B^+_c})(\epsilon_{D^{*+}_s}\cdot
p_{B^+_c})\frac{2A_2^{B^+_c\rightarrow
J/\psi}(m^2_{D^{*+}_s})}{m_{B^+_c}+m_{J/\psi}}](a_1V^*_{cb}V_{cs}+a_4V^*_{tb}V_{ts})\nonumber\\
&&+f_{J/\psi}m_{J/\psi}[(\epsilon_{J/\psi}\cdot\epsilon_{D^{*+}_s})(m_{B^+_c}+m_{D^{*+}_s})
A_1^{B^+_c\rightarrow D^{*+}_s}(m^2_{J/\psi})\nonumber\\
&&-(\epsilon_{J/\psi}\cdot p_{B^+_c})(\epsilon_{D^{*+}_s}\cdot
p_{B^+_c})\frac{2A_2^{B^+_c\rightarrow
D^{*+}_s}(m^2_{J/\psi})}{m_{B^+_c}+m_{D^{*+}_s}}]a_2V^*_{cb}V_{cs}\}\nonumber\\
M(B^+_c\rightarrow
D^{*+}_s\phi)&=&i\frac{G_F}{\sqrt{2}}[(\epsilon_{D^{*+}_s}\cdot\epsilon_{\phi})(m_{B^+_c}+m_{D^{*+}_s})A_1^{B^+_c\rightarrow
D^{*+}_s}(m^2_\phi)\nonumber\\
&&-(\epsilon_{\phi}\cdot p_{B^+_c})(\epsilon_{D^{*+}_s}\cdot
p_{B^+_c})\frac{2A_2^{B^+_c\rightarrow
D^{*+}_s}(m^2_\phi)}{m_{B^+_c}+m_{D^{*+}_s}}]a_4V^*_{tb}V_{ts}\nonumber\\
M(B^+_c\rightarrow D^+_s\phi)&=&i\sqrt{2}G_Ff_\phi
m_\phi(\epsilon_\phi\cdot p_B)F_1^{B^+_c\rightarrow
D^+_s}(m^2_\phi)a_4V^*_{tb}V_{ts}\nonumber\\
M(B^+_c\rightarrow
D^+K^0)&=&i\frac{G_F}{\sqrt{2}}f_{K^0}(m^2_{B^+_c}-m^2_{D^+})F^{B^+_c\rightarrow
D^+}
(m^2_{K^0})a_4V^*_{tb}V_{ts}\nonumber\\
M(B^+_c\rightarrow
D^+_sP)&=&i\frac{G_F}{\sqrt{2}}f_P(m^2_{B^+_c}-m^2_{D^+_s})F^{B^+_c\rightarrow
D^+_s} (m^2_P)(a_2V^*_{ub}V_{us}+a_3V^*_{tb}V_{ts}),
\end{eqnarray}}
with $P=\pi^0, \rho^0, \omega, \eta$; $a_2=c_2+c_1/3$ and
$a_3=c_3+c_4/3$.

\section{Amplitudes of the long distance processes}

\begin{figure}[t]
\centering \includegraphics[scale=0.7]{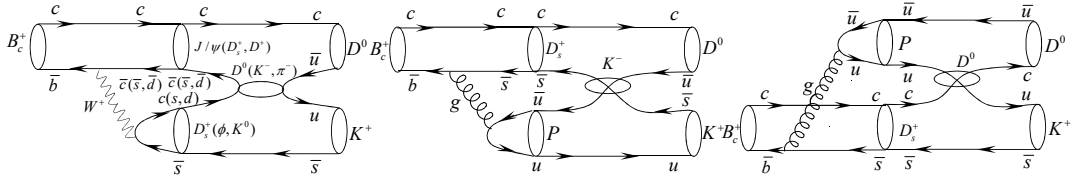}
\caption{\label{fig:b3}Diagrams of meson vertexes.}
\end{figure}

\begin{figure}[t]
\centering \includegraphics[scale=0.7]{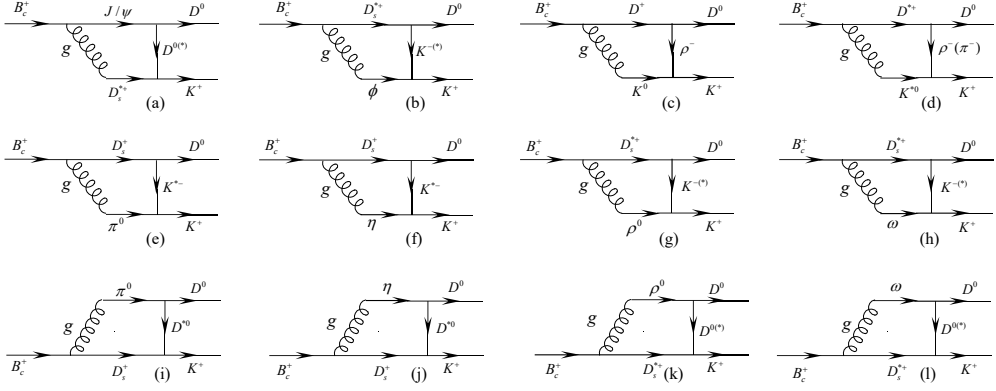}
\caption{\label{fig:b4}Diagrams of meson vertexes.}
\end{figure}

We know the above mentioned decays as intermediate state decays.
In this process, first the decaying meson decays into two
intermediate mesons, then, these two intermediate mesons are
changed into final mesons by exchanging a meson. FSI is done using
two channels named T-channel and cross section-channel. In
T-channel, the two final mesons share co-flavour quark and
anti-quark with their same flavour. But, in cross section channel,
two final mesons exchange two non-flavoured quarks or anti-quarks,
crosswisely. As we know, the $D^0$ and $K^+$ mesons constructed
from $c\bar{u}$ and $u\bar{s}$ quark anti-quark, respectively.
Then, these two mesons, which are in the final state, can share
quark anti-quark with their same flavour i.e. u quark in channel
T. On the other hand, intermediate state mesons can be produced by
sharing co-flavoured quark anti-quark like c, s and d which can be
$J/\psi D_s^{+(*)}$, $D_s^{+(*)}\phi$ and $D^{+(*)}K^{0(*)}$
mesons. Interchanging mesons of these processes are called $D^0$,
$K^-$ and $\pi^-$, respectively. In cross section channel, two
final state mesons $D^0$ and $K^+$ exchange $\bar{u}$ and
$\bar{s}$ anti-quarks, respectively. In this state, $D^+_sP$ with
$P=\omega$, $\rho^0$, $\eta$ and $\pi^0$ are intermediate mesons
and $K^-$ is interchange meson. In this channel, there is another
state in which two final mesons $D^0$ and $K^+$ mesons exchange c
and u quarks, respectively. Thus, intermediate mesons are the same
as the previous i.e. $D^+_sp$ and $D^0$ is the intermediate meson.
Fig. \ref{fig:b3}
illustrates FSI diagrams for $B^+_c\rightarrow D^0K^+$ decay
through T and cross section channels. The form factors
corresponded to FSI are different from those defined in the
previous section. In the two-body decay of previous section the
exchanged particle in interaction i.e. propagator is the
elementary particle boson or gluon, but in FSI the exchange
particle is a meson. Since, the form factor depends on mass and
momentum of particle, here too, we introduce the FSI form factor
as a function of mass and momentum
$F(q^2,m_i^2)=(\Lambda^2-m_i^2)/(\Lambda^2-q^2)$ in which $m_i$
and q are the mass and momentum parameter that shows the
effectiveness of strong interaction in a weak interaction through
$\Lambda=m_i+\eta\Lambda_{QCD}$. The parameter $\Lambda_{QCD}$ is
strong interaction energy scale which has the range of from 225
MeV to 750 MeV. We usually consider the value of $\Lambda_{QCD}$
constant equal to 225 MeV. Then, we change the phenomenological
parameter $\eta$. The range of this parameter is defined according
to the exchanged mesons. In Ref. \cite{X.Li1}, D and $D^*$ are
exchanged mesons which the authors have considered the range of
$\eta$ parameter from 0.5 to 3. However, in Ref \cite{P.Co}, for
similar exchanged mesons, authors have considered 5 for parameter
$\eta$. In Ref \cite{C.Me}, calculations are made for $\eta=4$. In
present paper, we have considered the range of $\eta$ from 1 to 3.
The more the value of $\eta$, the more the effect of strong
interaction. The mesons vertex factor is another important factor
in FSI. This factor is proportional to the coupling constant of
meson in vertices. There are three mesons in the top vertex and
three mesons in the down vertex which should follow meson vertex
rules. The first of them says that there must be at least one
vector meson in each vertex. Also, vector mesons should be
symmetric in the final and intermediate state in the manner that
if two final mesons are pseudo scalar, the intermediate mesons
both should either be pseudo scalar or vector mesons. In the decay
considered in this article, $B^+_c\rightarrow D^0K^+$, because two
final mesons are pseudo scalar, the intermediate mesons both
should be either pseudo scalar or vector mesons. In the case in
which both intermediate mesons are pseudo scalar, the exchanged
meson should be vector meson. In the case in which both
intermediate mesons are vector , the intermediate mesons can be
both scalar and vector. Applying these rules, the following decays
in the T-channel can be calculated: {\setlength\arraycolsep{.75pt}
\begin{eqnarray}
B^+_c&\rightarrow& J/\psi D^{*+}_s\rightarrow D^0K^+\;\;
\rm{exchang\;mesons\;are}\;D^0, D^{*0}\nonumber\\
B^+_c&\rightarrow& D^+K^0\rightarrow D^0K^+\;\;
\rm{exchang\;meson\;is}\;\rho^-\nonumber\\
B^+_c&\rightarrow& D^{*+}K^{*0}\rightarrow D^0K^+\;\;
\rm{exchang\;mesons\;are}\;\pi^-,\rho^-\nonumber\\
B^+_c&\rightarrow& D^{*+}_s\phi\rightarrow D^0K^+\;\;
\rm{exchang\;mesons\;are}\;K^-, K^{*-}.
\end{eqnarray}}
Also in the cross section channel the following decays are
calculated: {\setlength\arraycolsep{.75pt}
\begin{eqnarray}
B^+_c&\rightarrow& D^+_s\pi^0\rightarrow D^0K^+\;\;
\rm{exchang\;meson\;is}\;D^{*0}\nonumber\\
B^+_c&\rightarrow& D^+_s\eta\rightarrow D^0K^+\;\;
\rm{exchang\;meson\;is}\;D^{*0}\nonumber\\
B^+_c&\rightarrow& D^{*+}_s\rho^0\rightarrow D^0K^+\;\;
\rm{exchang\;mesons\;are}\;D^0, D^{*0}\nonumber\\
B^+_c&\rightarrow& D^{*+}_s\omega\rightarrow D^0K^+\;\;
\rm{exchang\;mesons\;are}\;D^0, D^{*0}\nonumber\\
B^+_c&\rightarrow& D^+_s\pi^0\rightarrow D^0K^+\;\;
\rm{exchang\;meson\;is}\;K^{*-}\nonumber\\
B^+_c&\rightarrow& D^+_s\eta\rightarrow D^0K^+\;\;
\rm{exchang\;meson\;is}\;K^{*-}\nonumber\\
B^+_c&\rightarrow& D^{*+}_s\rho^0\rightarrow D^0K^+\;\;
\rm{exchang\;mesons\;are}\;K^-, K^{*-}\nonumber\\
B^+_c&\rightarrow& D^{*+}_s\omega\rightarrow D^0K^+\;\;
\rm{exchang\;mesons\;are}\;K^-, K^{*-}.
\end{eqnarray}}
The meson vertices corresponded to FSI are seen in the Fig.
\ref{fig:b4}.
For example, consider to the meson vertex $\phi KK$. This vertex
shows the decay of $\phi$ meson into two KK mesons. The coupling
constant of the vertex is obtained from the relation $g_{\phi
KK}=(m_\phi/|\overrightarrow{p}_K|)\sqrt{(6\pi\Gamma^{exp}_{\phi\rightarrow
KK})/|\overrightarrow{p}_K|}$ in which in the particles data group
(PDG), $\Gamma^{exp}_{\phi\rightarrow KK}$ is given 2.09 MeV
\cite{M.Ta}. The parameter $|\vec{p}|$ is the K-meson's momentum
in the $\phi$-meson's rest frame. The rest of the coupling
constant of their meson vertices, can be obtained from the similar
relation for $\phi$-meson decay. The vertex factors which include
a vector meson V and two pseudo scalar mesons P, can be obtained
from $\langle
P_1(p_1)P_2(p_2)|i\ell|V_3(\epsilon_3,p_3)\rangle=-ig_{p_1p_2V}\epsilon_3\cdot(p_1+p_2)$.
As an example, vertex factor of $J/\psi DD$ and $\phi KK$ are:
{\setlength\arraycolsep{.75pt}
\begin{eqnarray}
\langle
D(p_1)D(p_2)|i\ell|J/\psi(\epsilon_3,p_3)\rangle&=&-ig_{J/\psi
DD}\epsilon_3\cdot(p_1+p_2),\nonumber\\
\langle K(p_1)K(p_2)|i\ell|\phi(\epsilon_3,p_3)\rangle&=&-ig_{\phi
KK}\epsilon_3\cdot(p_1+p_2).
\end{eqnarray}}
The factor of the $\langle
P_1(p_1)V_2(\epsilon_2,p_2)|i\ell|V_3(\epsilon_3,p_3)\rangle=-i\sqrt{2}g_{p_1V_2V_3}\epsilon_{\mu\nu\alpha\beta}
\epsilon_2^\mu\epsilon_3^{*\nu }p_1^\alpha p_2^\beta$ is used for
the vertex factors which include a pseudo scalar meson P and two
vector mesons V, so, the vertex factor of $J/\psi D^*D$ and $\phi
K^*K$ are: {\setlength\arraycolsep{.75pt}
\begin{eqnarray}
\langle
D(p_1)D^*(\epsilon_2,p_2)|i\ell|J/\psi(\epsilon_3,p_3)\rangle&=&-i\sqrt{2}g_{J/\psi
D^*D}\epsilon_{\mu\nu\alpha\beta}\epsilon_2^\mu\epsilon_3^{*\nu} p_1^\alpha p_2^\beta\nonumber\\
\langle
K(p_1)K^*(\epsilon_2,p_2)|i\ell|\phi(\epsilon_3,p_3)\rangle&=&-i\sqrt{2}g_{\phi
K^*K}\epsilon_{\mu\nu\alpha\beta}\epsilon_2^\mu\epsilon_3^{*\nu}
p_1^\alpha p_2^\beta.
\end{eqnarray}}
Finally, one can calculate the amplitude of graphs that have meson
loops as shown in Fig. \ref{fig:b4}, for the case in which both
mesons are pseudo scalar as: {\setlength\arraycolsep{.75pt}
\begin{eqnarray}\label{eq:17}
M(B^+_c(p_{B^+_c})&\rightarrow& P_1(p_1)P_2(p_2)\rightarrow
P_3(p_3)P_4(p_4))=\frac{1}{2}\int\frac{d^3\overrightarrow{p_1}}{2E_1(2\pi)^3}\frac{d^3\overrightarrow{p_2}}{2E_2(2\pi)^3}\nonumber\\
&&\times (2\pi)^4 \delta^4(p_{B^+_c}-p_1-p_2)M(B^+_c\rightarrow
P_1P_2)G(P_1P_2\rightarrow P_3P_4),
\end{eqnarray}}
in Eq. (\ref{eq:17}), $P_1P_2$ and $P_3P_4$ mesons are
intermediate and final state mesons, respectively and
$G(P_1P_2\rightarrow P_3P_4)$ shows meson vortices factors which
include the product of top factor in down factor in each graph.
For the case which both intermediate mesons are vector mesons, the
amplitude of graph, including meson loop is obtained from the
following equation: {\setlength\arraycolsep{.75pt}
\begin{eqnarray}\label{eq18}
M(B^+_c(p_{B^+_c})&\rightarrow&
V_1(\epsilon_1,p_1)V_2(\epsilon_2,p_2)\rightarrow
P_3(p_3)P_4(p_4))=\frac{1}{2}\int\frac{d^3\overrightarrow{p_1}}{2E_1(2\pi)^3}\frac{d^3\overrightarrow{p_2}}{2E_2(2\pi)^3}\nonumber\\
&&\times (2\pi)^4
\delta^4(p_{B^+_c}-p_1-p_2)f_{V_1}m_{V_1}V_{CKM}[(\epsilon_1^*\cdot\epsilon_2^*)(m_{B^+_c}+m_2)A_1^{B^+_c\rightarrow
V_2}(m_1^2)\nonumber\\
&&-(\epsilon^*_1\cdot p_{B^+_c})(\epsilon^*_2\cdot
p_{B^+_c})\frac{2A_2^{B^+_c\rightarrow
V_2}(m_1^2)}{m_{B^+_c}+m_2}]G(P_1P_2\rightarrow P_3P_4),
\end{eqnarray}}
which is assumed in that the vector mesons $V_2$ and $V_1$ are set
in form factor and vacuum state, respectively. But, in the case
with vector mesons $V_1$ and $V_2$ are setted in form factor and
vacuum state, respectively, indices are replaced in Eq.
(\ref{eq18}). So, the first amplitude of the nineteen amplitudes
of Fig. \ref{fig:b4}. i.e amplitude $B^+_c\rightarrow D^0K^+$,
with $D^0(q)$ as an exchanged meson, can be written as:
{\setlength\arraycolsep{.75pt}
\begin{eqnarray}\label{eq14}
M(3a,D^0)&=&i\frac{G_F}{4\sqrt{2}\pi
m_{B^+_c}}f_{D^{*+}_s}m_{D^{*+}_s}a_1V^*_{cb}V_{cs}g_{J/\psi
DD}g_{D_s^*DK}\int^{+1}_{-1}|\overrightarrow{p}_1|d(cos\theta)\frac{F^2(q^2,m_D^2)}{q^2-m_D^2}\nonumber\\
&&\times[(m_{B^+_c}+m_{J/\psi})A_1^{B^+_c\rightarrow
J/\psi}(m^2_{D^{*+}_s})K_1-\frac{2A_2^{B^+_c\rightarrow
J/\psi}(m^2_{D^{*+}_s})}{m_{B^+_c}+m_{J/\psi}}K'_1],
\end{eqnarray}}
in Eq. (\ref{eq14}) $\theta$ is the angle between momentums
$\overrightarrow{p}_1$ and $\overrightarrow{p}_3$, also
$q=p_1-p_3=p_4-p_2$ is the momentum of exchanged meson. In this
case, $q^2-m_D^2$ has the form
$m_1^2-2E_1E_3+2|\overrightarrow{p}_1||\overrightarrow{p}_3|cos\theta$.
The parameters $K_1$ and $K'_1$ show the product of polarization
vectors of vector mesons where $K_1=(\epsilon_1\cdot
p_3)(\epsilon_2\cdot p_4)(\epsilon^*_1\cdot\epsilon_2^*)$ and
$K'_1=(\epsilon_1\cdot p_3)(\epsilon_2\cdot p_4)(\epsilon_1\cdot
p_2)(\epsilon_2\cdot p_1)$. The second amplitude shows the same
process as before with the difference that the exchanged meson in
this case is vector meson, so, we have:
{\setlength\arraycolsep{.75pt}
\begin{eqnarray}
M(3a,D^{*0})&=&i\frac{G_F}{8\sqrt{2}\pi
m_{B^+_c}}f_{D^{*+}_s}m_{D^{*+}_s}a_1V^*_{cb}V_{cs}g_{J/\psi
D^*D}g_{D_s^*D^*K}\int^{+1}_{-1}|\overrightarrow{p}_1|d(cos\theta)\frac{F^2(q^2,m_{D^*}^2)}{q^2-m_{D^*}^2}\nonumber\\
&&\times[(m_{B^+_c}+m_{J/\psi})A_1^{B^+_c\rightarrow
J/\psi}(m^2_{D^{*+}_s})K_2-\frac{2A_2^{B^+_c\rightarrow
J/\psi}(m^2_{D^{*+}_s})}{m_{B^+_c}+m_{J/\psi}}K'_2],
\end{eqnarray}}
where
$K_2=\epsilon_{\mu\nu\alpha\beta}\epsilon_1^\mu\epsilon_{D^*}^\nu
p_3^\alpha
p_1^\beta\epsilon_{\rho\sigma\lambda\eta}\epsilon_2^\rho\epsilon_{D^*}^\sigma
p_4^\lambda p_2^\eta (\epsilon_1^*\cdot\epsilon_2^*)$ and
$K'_2=\epsilon_{\mu\nu\alpha\beta}\epsilon_1^\mu\epsilon_{D^*}^\nu
p_3^\alpha
p_1^\beta\epsilon_{\rho\sigma\lambda\eta}\epsilon_2^\rho\epsilon_{D^*}^\sigma
p_4^\lambda p_2^\eta (\epsilon_1\cdot p_{B_c})(\epsilon_2\cdot
p_{B_c})$. In this amplitude, $q^2-m_{D^*}^2$ has the form of
$m_1^2+m_3^2-m_{D^*}^2-2E_1E_3+2|\overrightarrow{p}_1||\overrightarrow{p}_3|cos\theta$.
{\setlength\arraycolsep{.75pt}
\begin{eqnarray}
M(3b,K^-)&=&i\frac{G_F}{4\sqrt{2}\pi m_{B^+_c}}f_\phi m_\phi
a_4V^*_{tb}V_{ts}g_{D^*_sDK}g_{\phi KK}\int^{+1}_{-1}|\overrightarrow{p}_1|d(cos\theta)\frac{F^2(q^2,m_K^2)}{q^2-m_K^2}\nonumber\\
&&\times[(m_{B^+_c}+m_{D^{*+}_s})A_1^{B^+_c\rightarrow
D^{*+}_s}(m^2_\phi)K_1-\frac{2A_2^{B^+_c\rightarrow
D^{*+}_s}(m^2_\phi)}{m_{B^+_c}+m_{D^{*+}_s}}K'_1].
\end{eqnarray}}
To calculate the amplitude $M(3b,K^{*-})$ it is enough to do as
follow:\\
1) convert 4 to 8 in the denominator of the first fraction line,\\
2) replace $g_{D^*_sDK}g_{KK\phi}$ with
$g_{D^*_sDK^*}g_{KK^*\phi}$,\\
3) replace $m_K$ with $m_{K^*}$ in the face and denominator of the
second fraction line,\\
4) convert coefficients $K_1$ and $K'_1$ to $K_2$ and $K'_2$,
respectively.\\
The FSI amplitude of the third graph of figure 3, the amplitude
$B^+_c\rightarrow D^+(p_1)K^0(p_2)\rightarrow D^0(p_3)K^+(p_4)$
with $\rho^-(q)$ as exchange meson can be written as:
{\setlength\arraycolsep{.75pt}
\begin{eqnarray}
M(3c)&=&-g_{DD\rho}g_{KK\rho}\int^{+1}_{-1}\frac{|\overrightarrow{p}_1|d(cos\theta)}{16\pi
m_{B^+_c}}M(B^+_c\rightarrow
D^+K^0)\frac{F^2(q^2,m_\rho^2)}{q^2-m_\rho^2}K_3,
\end{eqnarray}}
where $K_3= p_{1\mu}\epsilon_\rho^\mu p_{2\nu}\epsilon_\rho^\nu$.
The amplitude of the remaining graphs of figure 3 are obtained as
written amplitudes. Finally, the total amplitude of the FSI for
$B^+_c\rightarrow D^0K^+$ decay is calculated as:
{\setlength\arraycolsep{.75pt}
\begin{eqnarray}
M_{FSI}(B^+_c\rightarrow
D^0K^+)&=&M(3a,D)+M(3a,D^*)+M(3b,K)+M(3b,K^*)\nonumber\\
&&+M(3c)+M(3d,\pi)+M(3d,\rho)\ldots M(3l,K)+M(3l,K^*).
\end{eqnarray}}
At the end, one can calculate the branching ratio as:
{\setlength\arraycolsep{.75pt}
\begin{eqnarray}
\mathcal{B}(B^+_c\rightarrow
D^0K^+)&=&\frac{1}{\Gamma_{{\rm{tot}}}}\frac{|M(B^+_c\rightarrow
D^0K^+)|^2}{16\pi m_{B^+_c}},
\end{eqnarray}}
where $\Gamma_{\rm{tot}}=4.219\times10^{-13}$ GeV \cite{Z.Xi}.

\section{Experimental and theoretical dada}

The meson masses and decay constants needed in our calculations
are listed in the section (I) of the table \ref{tab1}. The
Cabibbo-Kobayashi-Maskawa (CKM) matrix is a $3\times3$ unitary
matrix, the elements of this matrix can be parameterized by three
mixing angles $A$, $\lambda$, $\rho$ and a CP-violating phase
$\eta$ \cite{M.Ta}: $V_{us}=\lambda$,
$V_{ub}=A\lambda^3(\rho-i\eta)$, $V_{cs}=1-\lambda^2/2$,
$V_{cb}=A\lambda^2$, $V_{ts}=-A\lambda^2$ and $V_{tb}=1$, the
results are shown in section (II) of the table \ref{tab1}. The
values of the coupling constants and the corresponding form
factors are given in sections (III) and (IV) of the table
\ref{tab1}, respectively. The Wilson coefficients $c_1$, $c_2$,
$c_3$ and $c_4$ in the effective weak Hamiltonian have been
reliably evaluated by the next-to-leading logarithmic order. To
proceed, we use the following numerical values at $\mu=m_b$ scale,
which have been obtained in the NDR scheme, these coefficient
numbers are inserted in the section (V) of the table \ref{tab1}
\cite{M.Be1}.
\begin{table}[t]
\centering\caption{\label{tab1}Input parameters.}
\begin{tabular}{c c c c c c c c}
  I) Mesons masses &\\ decay
  constants &\\ (in units of MeV) \cite{M.Ta}&\\
  \hline
  $m_{B^+_c}=6274.9\pm0.08$ & $m_{J/\psi}=3096.900\pm0.006$ & $m_{D^{+*}_s}=2112.2\pm0.4$ &\\
  $m_{D^+_s}=1968.34\pm0.07$ & $m_{D^0}=1864.83\pm0.05$ & $m_\phi=1019.461\pm0.016$ &\\
  $m_\omega=782.65\pm0.12$ & $m_{\rho^0}=775.26\pm0.25$ & $m_\eta=547.862\pm0.017$ &\\
  $m_{K^0}=497.611\pm0.013$ & $m_{K^+}=493.677\pm0.016$ & $m_{\pi^0}=134.9773\pm0.0005$ &\\
  $f_{B_c}=489\pm4$& $f_{J/\psi}=418\pm9$ & $f_{D^*_s}=315\pm8$ &\\
  $f_{D_s}=294\pm7$ & $f_D=234\pm15$ & $f_\phi=215\pm5$ &\\
  $f_\rho=210\pm4$ & $f_\omega=195\pm2$ & $f_K=159.8\pm1.84$ &\\
  $f_\pi=130.70\pm0.46$ & $f_\eta=63.6\pm0.23$ & \\
  \hline\hline
  \rule{0pt}{5ex}II) CKM matrix elements \cite{M.Ta}&\\
  \hline
  $V_{ub}=0.00394\pm0.00036$ & $V_{cb}=0.0422\pm0.0008$ & $V_{tb}=1.019\pm0.025$ &\\
  $V_{us}=0.2243\pm0.0005$ & $V_{cs}=0.997\pm0.017$ & $V_{ts}=0.0394\pm0.0023$ &\\
  \hline\hline
  \rule{0pt}{5ex}III) Coupling constants&\\
  \hline
  $g_{DD\rho}=2.52$ & $g_{D^*D\rho}=2.82$ & $g_{J/\psi DD}=7.71$ &\\
  $g_{J/\psi D^*D}=8.64$ \cite{Y.S.O} & $g_{D^*D\pi}=12.5$ \cite{V.M.B} & $g_{D^*_sDK}=18.34$ \cite{H.Y.C} &  &\\
  $g_{D^*_sD^*K}=9.23$ & $g_{K^*K\pi}=4.6$ \cite{C.D.L} &\\
  $g_{K^*K\rho}=g_{K^*K\phi}=6.48$ & $g_{KK\rho}=g_{KK\phi}=5.55$ \cite{X.Li2} &\\
  \hline\hline
  \rule{0pt}{5ex}IV) Form factors \cite{R.Dh}&\\
  \hline
  $A_0^{B_c\rightarrow J/\psi}=0.58^{+0.01}_{-0.03}$ & $A_1^{B_c\rightarrow J/\psi}=0.63^{+0.03}_{-0.03}$ &
  $A_2^{B_c\rightarrow J/\psi}=0.74^{+0.05}_{-0.06}$ &\\
  $A_1^{B_c\rightarrow D^{+*}_s}=0.18^{+0.01}_{-0.02}$ & $A_2^{B_c\rightarrow D^{+*}_s}=0.20^{+0.02}_{-0.03}$ &\\
  $F^{B_c\rightarrow D}=0.057^{+0.06}_{-0.08}$ & $F^{B_c\rightarrow D_s}=0.15\pm0.01$ &\\
  \hline\hline
  \rule{0pt}{5ex}V) Wilson coefficients \cite{M.Be1}&\\
  \hline
  $c_1=1.081$ & $c_2=-0.190$ & $c_3=0.014$ &\\
  $c_4=-0.036$ &\\
  \hline\hline
 \end{tabular}
\end{table}
Using the parameters relevant for the $B^+_c\rightarrow D^0K^+$
decay, we get flavor averaged branching ratio for the QCD
factorization method as:
\begin{eqnarray}
\mathcal{B}_{QCDF}(B^+_c\rightarrow D^0K^+)=1.41\times10^{-7}.
\end{eqnarray}
Now, applying the effects of the FSI, we obtain the branching
ratios of $B^+_c\rightarrow D^0K^+$ decay with different values of
$\eta$, for $\eta=1$ we get the number of the
$(1.17\pm0.19)\times10^{-5}$, for $\eta=2$ the value is
$(5.89\pm0.92)\times10^{-5}$ and the result of our calculation is
$(11.65\pm1.73)\times10^{-5}$ by choosing $\eta=3$. The results
show that the branching ratios are very sensitive to the variation
of $\eta$.

\section{Conclusion}

The decay of $B^+_c\rightarrow D^0K^+$ was calculated
theoretically before being experimentally observed and before any
experimented data is measured for it. The result was
$1.34\times10^{-7}$ using QCD factorization approach and
$6.60\times10^{-5}$ using perturbative QCD method. Until the LHCb
collaboration obtained the experimental branching ratio of this
decay between $3.72\times10^{-5}$ and $11.16\times10^{-5}$. In
this work, we have calculated the contribution of the FSI, i.e.
inelastic rescattering processes to the branching ratio of
$B^+_c\rightarrow D^0K^+$ decay and find that it spans a
relatively wider range of $(1.17\sim11.65)\times10^{-5}$ which is
obviously larger than the theoretically predicted value and
comparable with the perturbative QCD prediction. Our predicted
also covers the experimental range.

\end{document}